\documentclass[titlepage,aps,showpacs,pra,
]{revtex4}

\usepackage{bm}
\usepackage{amssymb}
\usepackage{amsmath}
\usepackage{graphicx}
\usepackage{color}

\fontsize{10}{13}
\selectfont

\begin{document}
\fontsize{10}{13}
\selectfont
\title{Quantum logical gates operating on stored light}
\author{K. \surname{S\l owik}} 
\email{karolina@fizyka.umk.pl}
\author{A. Raczy\'nski} \author{J. Zaremba}
\affiliation{Faculty of Physics, Astronomy and Informatics,
Nicolaus Copernicus University, ul. Grudzi\c{a}dzka 5/7, 87-100 
Toru\'n, Poland,}
\author{S. Zieli\'nska-Kaniasty}
\affiliation{Institute of Mathematics and Physics, 
University of Technology and Life Sciences, Al. Kaliskiego 7, 85-798
Bydgoszcz, Poland.}

\begin{abstract}
An implementation is proposed of single qubit gates, e.g., phase, NOT, $\sqrt{\mathrm{NOT}}$ and Hadamard, operating on polarized photons and based on light storage. Instead of processing photons themselves, qubit transformations are performed on atomic excitations due to photon storage in a medium of atoms in the tripod configuration.
\end{abstract}

\maketitle

Rapidly developing quantum information science requires novel implementations of quantum logical elements, in particular carriers of qubits and logical gates to process information. Photons are very efficient and comfortable carriers of information, robust with respect to decoherence. Processing them requires however an atomic medium through which they could effectively interact. Various optical implementations of logical operations on photon-carried qubits, based on different physical phenomena have been proposed \cite{kok}, of which an important class take advantage of the electromagnetically induced transparency (EIT) (\cite{harris}, for a review see, e.g., Ref. \cite{marangos}). It has been shown that it is possible to build phase gates using resonantly enhanced Kerr effect for weak pulses \cite{rebic,joshi, friedler} or SWAP gates with two photons interacting via an optically dressed medium in a particular double-lambda configuration \cite{yavuz}. A special realization of EIT leads to "stopping" light; in that case  information carried by probe photons is stored in the form of coherent atomic excitations (\cite{liu, philips, Fleischhauer}, for a review see, e.g., Ref. \cite{EIT}). The generic system to perform light storage is a three-level atomic lambda system, but to store a photon which may be in two possible basic states (i.e. a qubit), and then to process the information stored, one needs richer schemes admitting more active atomic levels and laser couplings. Such an interesting and flexible model may be the tripod system \cite{mazets, li}, combining two lambda systems which share a common coupling. In particular it allows for a controlled propagation/storage of a probe field under a control of two driving ones or a propagation of two probe fields controlled by a single driving one. The presence of three long-living lower states gives an opportunity to create two different atomic excitations at the storage stage and opens a way to processing the information supplied by the stopped photon. One can thus realize logical operations on photon states through (i) storing photons, (ii) performing the required operations on atomic states instead of photons themselves by applying pulses of additional electromagnetic fields and (iii) releasing photons in a transformed state. Wang \textit{et al.} \cite{wang} presented an idea of obtaining time-entangled photon states (time bins) by combining light storage with the fractional STIRAP technique. In our earlier papers we have shown that an optically dressed medium of atoms in the tripod configuration can serve as a beam splitter in time domain \cite{polariton, HOM}, working on stored light. 

Note that atomic media in the lambda configuration or its extensions, like double - lambda or tripod, have also been used for constructing logical gates in physical realizations other than slow or stored light and/or with qubit carriers other than photons. For example, logical operations based on STIRAP techniques performed on tripod states as qubit states were proposed in Refs. \cite{kis, moller1}. In Ref. \cite{koshino} it was first shown how to process qubits carried by both polarized photons and atoms in a cavity and then how to turn atom-photon gates into deterministic photon-photon $\sqrt{\mathrm{SWAP}}$ gates. 

Below we show how it is possible to implement any photon-carried qubit gate, but operating on stored light in an atomic medium in the tripod configuration.  It is well known that such gates are elementary components of quantum circuits \cite{nielsen}. In our realization the basic qubit states are photon circular polarization states $\sigma^+$ and $\sigma^-$. A photon carrying a quantum information propagates inside the medium in the EIT conditions due to a single $\pi$-polarized control field. Switching the latter off leads to photon storage in the form of atomic excitations. An effective two-photon interaction between two lower atomic states provides a transformation performed on the excitations. Switching the control field on again leads to light release with the photon in the transformed polarization state. 

We consider a five-level system (a tripod with three long-living lower states plus an additional upper state, see Fig. 1a), which can be physically realized in a cold medium of $^{87}$Rb atoms, with three lower levels $b$, $c$ and $b'$ corresponding to Zeeman sublevels of the state $|5S_{\frac{1}{2}}, F=1>$ with the magnetic number $M=-1,0,+1$ respectively, an upper level $a$ corresponding to $|5P_{\frac{3}{2}},F=0>$ and another level $f$ which can be chosen as $|5P_{\frac{3}{2}},F=1,M=0>$. Let $E_j$ denote the energy of the $j$-th level.

A strong classical field $\mathcal{E}_c = \epsilon_c(t) e^{i(k_c x-\omega_ct)} + c.c.$, propagating along the $x$ axis and polarized along the $z$ axis, couples the states $|a>$ and $|c>$ and makes the medium transparent \cite{EIT} for a probe photon which propagates along the $z$ axis. The state of the latter is in general a superposition of right and left circular polarization states which correspond to the basic qubit states and will be denoted by $+$ and $-$, respectively. The right and left circular polarization components couple the state $|a>$ with $|b>$ and $|b'>$, respectively, and are given by $\mathcal{E}_p^\pm = \epsilon_\pm(z,t) e^{i(k z-\omega t)} + h.c.$, where the wave packet $\epsilon_{\pm}(z,t) e^{i(k z-\omega t)} = \sum_m \sqrt{\frac{\hbar \omega_m}{2 \epsilon_0 V}} a_m^{(\pm)}e^{i(k_m z-\omega_m t)}$, $\omega_m$ is the frequency of the $m$-th mode of the quantum field, $\epsilon_0$ is the vacuum permittivity, $V$ is the quantization volume, $a_m^{(\pm)}$ is the annihilation operator of the $m$-th mode of right or left circular polarization, respectively, $\omega$ is the central frequency of the field and $k$ is its wave number.

The medium of atoms distributed in a continuous way \cite{Fleischhauer} is described by a set of flip operators: \\ $\sigma_{ba}(z) = |b><a|\exp{(i(\omega t-kz))}$, $\sigma_{ca}(z) = |c><a|\exp{(i(\omega_c t-k_cx))}$, \\$\sigma_{bc}(z) = |b><c|\exp{(i(\omega t-kz)-i(\omega_c t - k_c x))}$, etc., being slowly varying functions of $z$ and $t$ and fulfiling the commutation relation $[\sigma_{ij}(z),\sigma_{kl}(z')] = \delta(z-z') \frac{L}{N} (\delta_{jk}\sigma_{il}(z)-\delta_{il}\sigma_{kj}(z))$. 

The atom-field interaction hamiltonian in the rotating wave approximation reads 
\begin{equation}
H = -\frac{N_0}{L} \int_{0}^L dz (\epsilon_c d_{ac}\sigma_{ac} + \epsilon_+ d_{ab} \sigma_{ab} + \epsilon_- d_{ab'} \sigma_{ab'}) + h.c., 
\end{equation}
where $N_0$ denotes the number of atoms, $L$ is the sample length and $d_{ij}$ is the electric dipole moment matrix element.

The evolution of the medium is described by the equation $i\hbar \dot{\sigma} = [\sigma,H] + F$, where $F$ is the Langevin noise operator. The influence of the latter is however negligible \cite{Fleischhauer} and it will not be taken into account in our further considerations. In the resonance conditions, without dissipation, the evolution equations of the medium read in the first-order perturbation approximation with respect to the probe fields
\begin{eqnarray}
i\hbar \dot{\sigma}_{ba} &=& -\frac{1}{2} \Omega_+ - \Omega_c \sigma_{bc} - \Omega_- \sigma_{bb'}, \label{bloch1}\\
i\hbar \dot{\sigma}_{b'a} &=& -\frac{1}{2} \Omega_- - \Omega_c \sigma_{b'c} - \Omega_+ \sigma_{b'b},\\
i\hbar \dot{\sigma}_{bc} &=& -\Omega_c \sigma_{ba}, \\
i\hbar \dot{\sigma}_{b'c} &=& -\Omega_c \sigma_{b'a}, 
\end{eqnarray}
with the Rabi frequencies $\Omega_{\pm} = \frac{\epsilon_{\pm} d_{ab,ab'}}{\hbar}$, $\Omega_c = \Omega_c^* = \frac{\epsilon_c d_{ac}}{\hbar}$. The propagation equations for the two polarization modes of the probe field read in the adiabatic approximation (i.e. with $\dot{\sigma}_{ba}=\dot{\sigma}_{b'a} = 0$)
\begin{equation}
\left( \frac{\partial}{\partial t} + c \frac{\partial}{\partial z} \right) \Omega_{+,-} = i\kappa^2 \sigma_{ab,ab'} = -\kappa^2 \frac{1}{\Omega_c} \frac{\partial}{\partial t} \frac{1}{\Omega_c} \left( \frac{1}{2} \Omega_{+,-} + \sigma_{bb',b'b} \Omega_{-,+}\right), \label{prop0}
\end{equation}
with $\kappa^2 = \frac{N|d_{ab,ab'}|^2\omega}{2\epsilon_0 \hbar}$, $N$ being the number of atoms per unit volume. In the evolution equations some of the flip operators have been substituted by their initial mean values which do not change in the first order perturbation approximation, i.e. $\sigma_{aa} , \sigma_{cc} , \sigma_{ac} \approx 0$. We have additionally set $\sigma_{bb} = \sigma_{b'b'} = \frac{1}{2}$; these populations do not change either in this approximation. 

Note that the propagation equations are decoupled only if the medium is prepared in such a state that $\sigma_{bb'}=0$, which we will assume below. Then the propagation of each polarization mode is independent of the other. As we show below, in our protocols a nonzero $\sigma_{bb'}$ is never produced in the first order perturbation approximation with respect to the probe fields, due to equal populations of the states $|b>$ and $|b'>$.

Below we describe the protocols of manipulating the photon polarization given by $\overrightarrow{e} = \overrightarrow{e}_x \cos \frac{\vartheta}{2}e^{-i\frac{\varphi}{2}} + \overrightarrow{e}_y \sin \frac{\vartheta}{2} e^{i\frac{\varphi}{2}}$, which correspond to unitary transformations rotating a qubit vector on the Bloch sphere around the axes $X$, $Y$ and $Z$ ($X = \sin \vartheta \cos \varphi$, $Y = \sin \vartheta \sin \varphi$, $Z = \cos \vartheta$).

The processes of storing and releasing a photon can be conveniently described in the language of dark state polaritons \cite{Fleischhauer, polariton} - quasi-particles consisting of electromagnetic and atomic parts with annihilation operators given by
\begin{equation}
\psi_{+,-} = \sqrt{\frac{N_0}{\kappa^2 L}} \left( \Omega_{+,-} \cos \theta + \sqrt{2}\kappa \sigma_{bc,b'c} \sin \theta \right), \label{psi0}\\
\end{equation}
where $\tan \theta = \frac{\kappa}{\sqrt{2}\Omega_c}$. The dark-state polariton propagation equations follow from Eqs. (\ref{bloch1} - \ref{prop0}
) and read $\left( \frac{\partial}{\partial t} + c \cos^2 \theta (t) \frac{\partial}{\partial z} \right) \psi_{\pm}(z,t) = 0$, and have a simple solution $\psi_{\pm}(z,t) = \psi_{\pm}(z-\int_0^t c \cos^2 \theta (\tau) d\tau,0)$.

Initially the control field is strong, the mixing angle $\theta$ is small and both polaritons are almost purely electromagnetic (the probe photon propagates in the medium). When the control field is adiabatically turned off, the angle $\theta \rightarrow \frac{\pi}{2}$ and the polaritons become purely atomic. The $\psi_+$ ($\psi_-$) polariton describes the storage of the right (left) circularly polarized probe photon and the corresponding atomic excitation build-up, described by $\sigma_{bc}$ or $\sigma_{b'c}$, respectively. Such a process has no significant impact on the population distribution (it remains unchanged in the first order perturbation approximation with respect to the probe field). 

As our goal is to influence the polarization of the released photon, we need to manipulate the polariton at the storage stage, i.e. the atomic flip operators, as follows from Eqs. (\ref{psi0}
). To do that we can provide, e.g., a two-photon coupling between the states $|b>$ and $|b'>$ via an upper state $|f>$ of the magnetic number $M=0$ (see Fig. 1b). Such an interaction $V$ is then of the form $V = W|b><b'| + W^* |b'><b|$. The effective second - order coupling $W$ reads
\begin{equation}
W \equiv |W|e^{i\chi} = \frac{<b|U_+|f><f|U_-|b'>}{E_b+\hbar \omega_{U_+} - E_f},
\end{equation}
where $U_{\pm}$ are nonresonant laser couplings of frequencies $\omega_{U_{\pm}}$ and of right and left circular polarization, of the states $|b>$ and $|b'>$, respectively, with the state $|f>$. The frequencies $\omega_{U_{\pm}}$ are chosen to assure the two-photon Raman resonance $b-f-b'$, taking also into account the dynamical Stark shift of the levels $b$ and $b'$.
 
Due to the coupling $V$ the medium flip operators $\sigma_{ij}(t)$ evolve according to the equation $i\hbar \dot{\sigma}_{ij} = [\sigma_{ij},V]$, the solution of which reads $\sigma(\tau) = e^{\frac{i}{\hbar}V\tau}\sigma(\tau=0)e^{\frac{-i}{\hbar}V\tau}$ or, explicitly,
\begin{equation}
\sigma(\tau) = \left[ \begin{array}{ccc} \sigma_{bb} \cos^2 \frac{|W|\tau}{\hbar} + \sigma_{b'b'} \sin^2 \frac{|W|\tau}{\hbar} & \sigma_{bc} \cos \frac{|W|\tau}{\hbar} & -ie^{i\chi}\sin \frac{|W|\tau}{\hbar} \cos \frac{|W|\tau}{\hbar} (\sigma_{bb} - \sigma_{b'b'}) \\ 
\sigma_{cb} \cos \frac{|W|\tau}{\hbar} & \sigma_{cc} & -i e^{i\chi} \sigma_{cb} \sin \frac{|W|\tau}{\hbar} \\
ie^{-i\chi}\sin \frac{|W|\tau}{\hbar} \cos \frac{|W|\tau}{\hbar} (\sigma_{bb} - \sigma_{b'b'}) & i e^{-i\chi} \sigma_{bc} \sin \frac{|W|\tau}{\hbar} & \sigma_{bb} \sin^2 \frac{|W|\tau}{\hbar} + \sigma_{b'b'} \cos^2 \frac{|W|\tau}{\hbar} \end{array} \right] \label{y0}
\end{equation}
when the stored photon is right circularly polarized, and
\begin{equation}
\sigma(\tau) = \left[ \begin{array}{ccc} \sigma_{b'b'} \sin^2 \frac{|W|\tau}{\hbar} + \sigma_{bb} \cos^2 \frac{|W|\tau}{\hbar} & ie^{i\chi}\sigma_{b'c} \sin \frac{|W|\tau}{\hbar} & ie^{i\chi}\sin \frac{|W|\tau}{\hbar} \cos \frac{|W|\tau}{\hbar} (\sigma_{b'b'} - \sigma_{bb}) \\ 
-ie^{-i\chi}\sigma_{cb'} \sin \frac{|W|\tau}{\hbar} & \sigma_{cc} & \sigma_{cb'} \cos \frac{|W|\tau}{\hbar} \\
-ie^{-i\chi}\sin \frac{|W|\tau}{\hbar} \cos \frac{|W|\tau}{\hbar} (\sigma_{b'b'} - \sigma_{bb}) & \sigma_{b'c} \cos \frac{|W|\tau}{\hbar} & \sigma_{b'b'} \cos^2 \frac{|W|\tau}{\hbar} + \sigma_{bb} \sin^2 \frac{|W|\tau}{\hbar} \end{array} \right] \label{y1}
\end{equation}
when it is left circularly polarized. In the above equations $\tau$ is the duration of the interaction $W$ and the operators $\sigma_{ij}$ in the r.h.s. of Eqs. (\ref{y0}, \ref{y1}) correspond to $\tau=0$. In the assumed case of $\sigma_{bb} = \sigma_{b'b'} = \frac{1}{2}$ and $\sigma_{cc}=0$, the populations do not change in the first-order perturbation approximation, and thus no coherent excitation $\sigma_{bb'}$ is generated, so the two modes of the probe field remain uncoupled (see Eqs. (\ref{prop0}
)). 

From Eqs. (\ref{y0}, \ref{y1}) it follows that if a right (left) circularly polarized photon has been stopped, the excitation $\sigma_{bc}$ ($\sigma_{b'c}$) is modified and the initially zero $\sigma_{b'c}$ ($\sigma_{bc}$) is generated in the following way
\begin{equation}
\left[ \begin{array}{c} \sigma_{bc} \\ \sigma_{b'c} = 0 \end{array}\right] \rightarrow \left[ \begin{array}{c}\sigma_{bc} \cos \frac{|W|\tau}{\hbar} \\ i e^{-i\chi} \sigma_{bc} \sin \frac{|W|\tau}{\hbar} \end{array} \right], \,\,\,\,\,\,\, \left( \left[ \begin{array}{c} \sigma_{bc} = 0 \\ \sigma_{b'c} \end{array} \right] \rightarrow \left[ \begin{array}{c} ie^{i\chi} \sigma_{b'c} \sin \frac{|W|\tau}{\hbar} \\ \sigma_{b'c} \cos \frac{|W|\tau}{\hbar} \end{array} \right] \right).
\end{equation}

 In general, the stored photon has an arbitrary polarization, so the flip operators undergo a transformation given by $\left[ \begin{array}{c} \sigma_{bc} \\ \sigma_{b'c} \end{array} \right] \rightarrow G \left[ \begin{array}{c} \sigma_{bc} \\ \sigma_{b'c} \end{array} \right]$ with
\begin{equation}
G = \left[ \begin{array}{cc} \cos \frac{|W|\tau}{\hbar} & i e^{i\chi} \sin \frac{|W|\tau}{\hbar} \\ i e^{-i\chi} \sin \frac{|W|\tau}{\hbar} & \cos \frac{|W|\tau}{\hbar} \end{array} \right]. \label{gate} 
\end{equation}
The above considerations remain valid for any $|W|$ with a constant phase $\chi$ and smooth time-dependence provided that $|W|\tau$ is replaced by $\int_0^\tau |W(\tau')|d\tau'$.

The last stage of the protocol is the photon release. To perform it we adiabatically turn the control field $\Omega_c$ on. The $\theta$ angle decreases from $\frac{\pi}{2}$ to a small value, which corresponds to changing the polaritons from atomic to almost purely electromagnetic. Due to the one-to-one correspondence between the atomic excitations and the photon polarization states, the whole three-stage operation results in transforming $\left[ \begin{array}{c} \Omega_+ \\ \Omega_- \end{array} \right]$ into $G\left[ \begin{array}{c} \Omega_+ \\ \Omega_- \end{array} \right]$. Thus, the photon qubit is transformed as follows: $c_+ |+> + c_-|-> \rightarrow (G_{11}^* c_+ + G_{21}^* c_-) |+> + (G_{12}^* c_+ + G_{22}^* c_- ) |->$, with $\Omega_\pm^\dagger |vac> = |\pm>$. This means that one obtains a class of flexible logical gates defined by $G$ for photon polarization states. They correspond to rotations on the Bloch sphere and the final polarization of the photon is determined by controllable parameters, i.e. the phase $\chi$, which determines the rotation axis, and the time of interaction or the interaction strength, which determine the rotation angle. We give a few examples below. 

For $\chi = \pi$ the gate $G$ corresponds to a rotation around the $X$ axis about the angle $\beta = 2\frac{|W|t}{\hbar}$ 
\begin{equation}
G = e^{-i\frac{\beta}{2} \sigma_x} = R_X(\beta) = \left[ \begin{array}{cc} \cos \frac{\beta}{2} & -i\sin \frac{\beta}{2} \\ -i\sin \frac{\beta}{2} & \cos \frac{\beta}{2} \end{array} \right].
\end{equation}
By a proper choice of the interaction time and/or strength (i.e. for $\beta = \pi$) it is possible to obtain the NOT gate (multiplied by an overall phase factor $-i$), and for $\beta = \frac{\pi}{2}$ - the $\sqrt{\mathrm{NOT}}$ gate (multiplied by a phase factor $e^{-i\frac{\pi}{4}}$).
For $\chi = \frac{\pi}{2}$ we obtain a rotation around the $Y$ axis
\begin{equation}
G = e^{-i\frac{\beta}{2} \sigma_y} = R_Y(\beta) = \left[ \begin{array}{cc} \cos \frac{\beta}{2} & -\sin \frac{\beta}{2} \\ \sin \frac{\beta}{2} & \cos \frac{\beta}{2} \end{array} \right],
\end{equation}
which for $\beta = \pi$ gives the $\sigma_y$ Pauli matrix (multiplied by an overall phase $-i$), and for $\beta = \frac{\pi}{2}$, the gate $\tilde{H} = R_Y(\frac{\pi}{2}) = \frac{1}{\sqrt{2}}\left[ \begin{array}{cc} 1 & 1 \\ -1 & 1 \end{array} \right]$. The Hadamard gate, which is given by $H = \frac{1}{\sqrt{2}} \left[ \begin{array}{cc} 1 & 1 \\ 1 & -1 \end{array} \right]$, can be obtained as a combination of the rotations around $X$ and $Y$ axis (or $Y$ and $Z$ axis - see below), followed by an overall phase shift: $H = e^{i\frac{\pi}{2}} \tilde{H} R_X(\pi)= e^{i\frac{\pi}{2}} R_Z(\pi) \tilde{H}$.

To perform the qubit state rotation around the $Z$ axis, it is necessary to shift the phases of the excitations $\sigma_{bc}$ and $\sigma_{b'c}$. This can be done via the Zeeman effect \cite{mair, polariton}. The states $|b>$ and $|b'>$ correspond to Zeeman sublevels of opposite magnetic quantum numbers ($-1$ and $+1$ respectively). Therefore, when a magnetic field is applied at the storage stage, they are shifted in opposite directions, while the energy of the level $|c>$ of the magnetic number $0$ remains unchanged. Thus, the corresponding operators acquire phase shifts of exactly the same value but opposite signs, i.e. $\sigma_{bc} \rightarrow \sigma_{bc}e^{-i\frac{\phi}{2}}$, $\sigma_{b'c} \rightarrow \sigma_{b'c}e^{i\frac{\phi}{2}}$, $\frac{\phi}{2}$ being the pulse area of the magnetic interaction: $\frac{\phi}{2} = \int_0^\tau \frac{e g_F}{2m} B(\tau') d\tau'$, where $\tau$ is the interaction time, $g_F$ stands for the Land\'{e} g-factor and $B(\tau)$ is the value of the magnetic field induction at a given time $\tau$. Note that the acquired phase $\phi$ is easily controllable, as it is proportional to the interaction time and the applied magnetic field. 

The phase gate operation can be written in the matrix form
\begin{equation}
G = e^{-i\frac{\phi}{2} \sigma_{z}} = R_Z(\phi) = \left[ \begin{array}{cc} e^{-i\frac{\phi}{2}} & 0 \\ 0 & e^{i\frac{\phi}{2}} \end{array} \right].
\end{equation}

In conclusion, we have shown how to implement an arbitrary single qubit gate operating on a stored photon. A photon in a polarization state, being an arbitrary superposition of left and right circular polarization states, is stored in the corresponding combination of atomic excitations. An interaction coupling the atomic states allows for a controlled manipulation of the excitations. The released photon ends up in the polarization state determined by the latter. The incoming qubit has thus been transformed into an outgoing one in a controlled way. We have shown in particular how to implement arbitrary rotations on the Bloch sphere, in particular the NOT, $\sqrt{\mathrm{NOT}}$, Hadamard and phase gates.

\newpage
\begin{center}
\begin{figure}
\includegraphics[scale=0.6]{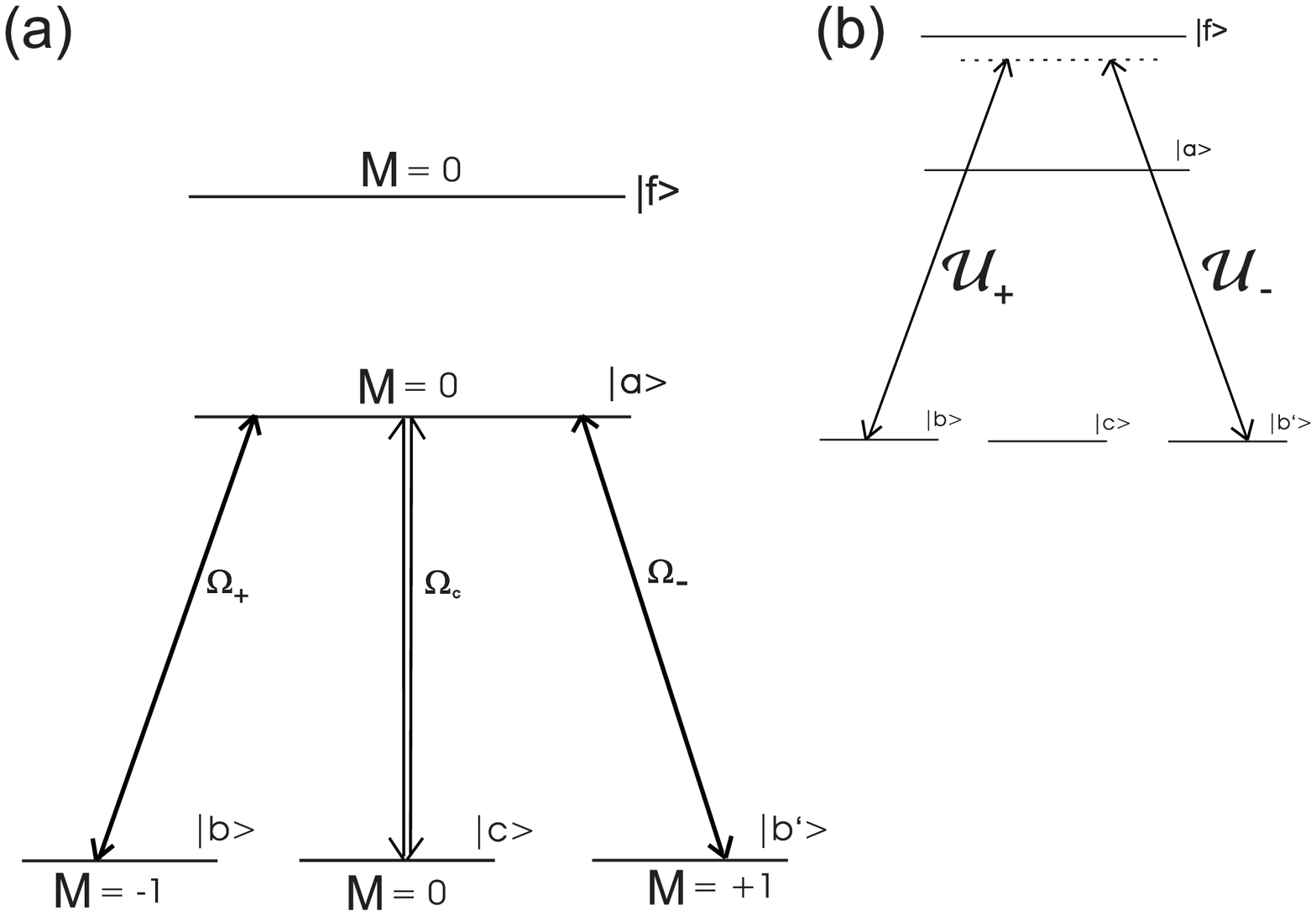}
\caption{Level and coupling diagram (a) elementary diagram, (b) auxiliary couplings at the storage stage.}
\end{figure}
\end{center}
\end{document}